# Neutral and Charged Inter-Valley Biexcitons in Monolayer MoSe$_2$

Kai Hao[1], Lixiang Xu[1], Judith F. Specht[2], Philipp Nagler[3], Kha Tran[1], Akshay Singh[1], Chandriker Kavir Dass[1], Christian Schüller[3], Tobias Korn[3], Marten Richter[2], Andreas Knorr[2], Xiaoqin Li[1*], and Galan Moody[4*]

[1] Department of Physics and Center for Complex Quantum Systems, University of Texas at Austin, Austin, TX 78712, USA.

[2] Institut für Theoretische Physik, Nichtlineare Optik und Quantenelektronik, Technische Universität Berlin, 10623 Berlin, Germany.

[3] Department of Physics, University of Regensburg, 93040 Regensburg, Germany.

[4] National Institute of Standards & Technology, Boulder, CO 80305, USA.

*e-mail: galan.moody@nist.gov, galan.moody@gmail.com; elaineli@physics.utexas.edu

**In atomically thin transition metal dichalcogenides (TMDs), reduced dielectric screening of the Coulomb interaction leads to strongly correlated many-body states, including excitons[1–5] and trions[6–10], that dominate the optical properties. Higher-order states, such as bound biexcitons, are possible but are difficult to identify unambiguously using linear optical spectroscopy methods alone. Here, we implement polarization-resolved two-dimensional coherent spectroscopy to unravel the complex optical response of monolayer MoSe$_2$ and identify multiple higher-order correlated states[11–14]. Decisive signatures of neutral and charged inter-valley biexcitons appear in cross-polarized two-dimensional spectra as distinct resonances with respective ~20 meV and ~5 meV binding energies—similar to recent calculations using variational and Monte Carlo methods[15,16]. A theoretical model taking into account the valley-dependent optical selection rules reveals the specific quantum pathways that give rise to these states. Inter-valley biexcitons identified here, comprised of neutral and charged excitons from different valleys, offer new opportunities for creating exotic exciton-polariton condensates[17] and for developing ultrathin biexciton lasers and polarization-entangled photon sources[18].**

The stable formation of tightly bound excitons and trions in atomically thin TMDs suggests that higher-order states comprised of correlations between four or more particles are possible (Fig. 1a). A bound biexciton, resulting from four-particle correlations between two electrons and two holes, is a hallmark of many-body interactions between quasiparticles in semiconductors. Experimental identification and microscopic calculations of biexcitons in conventional semiconductors and their heterostructures have greatly advanced our fundamental understanding of many-body physics in semiconductors[19]. A biexciton is a classic example for which a mean field (Hartree Fock) theoretical description clearly fails and higher-order electronic correlations have to be included[20–23].

Experimental and theoretical studies of biexcitons in monolayer TMDs present additional challenges compared to conventional semiconductors. The relatively large inhomogeneous broadening and possible spectral overlap between different resonances (biexcitons, trions, and excitons localized by defects) make it difficult to separate these states using the most widely used optical spectroscopy methods. In addition, microscopic calculations of biexcitons in 2D materials need to properly take into account anisotropic screening of the Coulomb interaction and the coupled electronic spin and valley pseudospin degrees of freedom. Using left (σ+) or right (σ-) circularly polarized light tuned close to the lowest-energy *A*-exciton resonance[24–26], excitons and trions are preferentially created at the *K* and *K'* valleys (Fig. 1b) and the valley index can be preserved during emission processes. Questions regarding the valley degree of freedom associated with higher-order bound states naturally arise. Do additional many-body states with different internal valley indices exist and exhibit distinct physical properties? While a few experimental studies have sought to investigate the bound biexciton in monolayer TMDs[27–30], significant discrepancies between experiment and theory[15,16] highlight a limited fundamental understanding of correlated states in atomically thin materials.

In this work, we shed new light on this topic by investigating monolayer MoSe$_2$ using resonant two-dimensional coherent spectroscopy (2DCS)[31]. This technique is highly sensitive to higher-order electron-hole correlations, which we leverage to distinguish between the exciton, trion, and different species of biexciton resonances. We show that a comparison of co- and cross-circularly polarized 2D spectra reveals previously unobserved resonances that are fully compatible with theoretically predicted neutral and charged biexciton states with binding energies of ~20 meV and ~5 meV (Refs. 15,16). Both types of biexciton states—spectrally isolated in the 2D frequency plane—are masked in conventional one-dimensional spectroscopy by the inhomogeneously broadened trion and exciton resonances. The polarization dependence of the 2D spectrum is reproduced with density matrix calculations taking into account the valley-dependent optical selection rules. These results suggest that both types of biexcitons are inter-valley in nature and consist of two excitons with large difference in crystal momentum, making them unique types of higher-order bound states with no direct analog in conventional semiconductors. The clear spectroscopy signatures of inter-valley biexcitons presented here provide guidance to microscopic theories of higher correlated states and raise interesting prospects for valley and photon-polarization entanglement in TMDs[18].

We examined monolayer MoSe$_2$ mechanically exfoliated onto a sapphire substrate (see Fig. 1c), which is held at a temperature of 20 K for optical spectroscopy experiments in transmission. A time-integrated photoluminescence spectrum acquired using continuous-wave 532 nm excitation is shown in Fig. 1d. The spectrum features two peaks at ~1650 meV and ~1620 meV that have been previously assigned to the exciton and negative trion, respectively[32]. The exciton is comprised of an electron-hole pair with opposite spins in the same valley, as illustrated in Fig. 1b. For the trion, the lowest-energy transition consists of an exciton in one valley bound to an additional electron with opposite spin in the opposite valley. Biexciton resonances are not isolated in the photoluminescence spectrum due to spectral overlap with the relatively broad inhomogeneous linewidths for the exciton and trion. In principle, the lowest energy bound biexciton transition comprises two excitons in opposite valleys, since the energy for bound biexcitons generated by simultaneous excitation of two excitons in the same valley is much higher[16,28]. Because the lowest-energy dipole-allowed transition in MoSe$_2$ is between the highest valence band and lowest conduction band, the possible biexciton, trion, and exciton configurations are simplified compared to tungsten-based TMDs[33].

To probe biexcitons in monolayer MoSe$_2$, we performed 2DCS experiments with carefully chosen excitation and polarization conditions. 2DCS is a three-pulse four-wave mixing (photon echo) technique with interferometric precision of the timing delays between the pulses and phase stabilization better than $\lambda/100$, where $\lambda$ is the excitation wavelength[34,35]. The laser spectrum is centered near 1620 meV, which is tuned below the exciton resonance energy to preferentially enhance any signatures of lower-energy bound biexcitons. The 2DCS experiments (details included in Methods section) are performed in the box geometry and with a rephasing pulse sequence (see Fig. 2a). Briefly, three sub-40 fs pulses with wavevectors $\mathbf{k}_1$, $\mathbf{k}_2$, and $\mathbf{k}_3$ interact nonlinearly with the sample to generate a four-wave mixing signal $S(t_1, t_2, t_3)$ that is detected in the phase-matched direction $\mathbf{k}_S = -\mathbf{k}_1 + \mathbf{k}_2 + \mathbf{k}_3$. We show the pulse time ordering in Fig. 2b. The four-wave mixing signal is heterodyne detected with a fourth phase-stabilized reference pulse and the resulting interference signal is spectrally resolved using a spectrometer. The signal is recorded as the delay between the first two excitation pulses, $t_1$, is scanned, while the delay between the second and third pulses, $t_2$, is held fixed. Fourier-transformation of the signal with respect to $t_1$ generates a 2D rephasing spectrum of the signal $S(\hbar\omega_1, t_2, \hbar\omega_3)$, which correlates the excitation energies ($\hbar\omega_1$) of the system during $t_1$ with its emission energies ($\hbar\omega_3$) during $t_3$. For all 2DCS experiments, the fluence of the excitation pulses is kept below ~4 µJ/cm² (~$10^{12}$ excitons/cm²), which is within the $\chi^{(3)}$ regime.

We first present a 2D amplitude spectrum in Fig. 3a acquired using co-circularly polarized excitation and detection, which is sensitive to excitonic transitions in only one valley. The spectrum features two peaks on the diagonal dashed line at ~1648 meV and ~1621 meV corresponding to excitation and emission from the exciton (X) and trion (T) transitions, respectively. The linewidths of the peaks along the diagonal ($\hbar\omega_1 = -\hbar\omega_3$) and cross-diagonal (perpendicular to the diagonal) directions reflect the inhomogeneous and homogeneous broadening of the exciton and trion resonances, which have been previously characterized[32,36]. The small Stokes shift of the peaks relative to the linear emission spectrum and the moderate inhomogeneous linewidths attest to the high quality of the material. The co-circular spectrum also features off-diagonal cross peaks at the excitation energy of the exciton and emission energy of the trion (XT) and vice-versa (TX), which are indicative of coherent interactions between the exciton and trion[10,32,37].

The 2D spectrum obtained using cross-circular polarization shown in Fig. 3b is the main result of this work. An additional peak (XX) appears in the spectrum red-shifted along the emission energy axis from the exciton, which we attribute to the bound biexciton state. The quantum pathway responsible for peak XX can be understood from the timing sequence in Fig. 2b as follows (see also the Liouville space pathways in the supplemental material): the first pulse (σ+) resonantly generates a coherence $g \leftrightarrow X$ between the crystal ground state ($g$) and the exciton state ($X$) in the K valley. After a delay $t_1$, the second pulse (σ-) drives the exciton transition in the K' valley. The Coulomb interaction between the two excitons induces an effective non-radiative coherence between excitons K and K' valleys during $t_2$. After a delay $t_2$, the third pulse (σ+) generates an inter-valley coherence $X \leftrightarrow XX$ between the exciton and biexciton states, which radiates as the four-wave mixing signal. This quantum pathway evolves during $t_1$ with energy $E_X$ and during $t_3$ with energy $E_X$-$\Delta_{XX}$, where $\Delta_{XX}$ is the neutral biexciton binding energy. In the Fourier domain, this contribution appears at the exciton excitation energy $E_X$ and emits red-shifted from the exciton peak by $\Delta_{XX}$, which is consistent with the bound biexciton peak *XX* with $\Delta_{XX} \approx 20$ meV. Because the state *XX* is only observable when the first two excitation pulses have opposite helicity, we conclude that this biexciton consists of two excitons with opposite valley index. While biexcitons have been identified using 2DCS in semiconductor quantum wells[11], quantum dots[38], and bulk layered materials[39], the observation of inter-valley biexcitons consisting of two excitons in opposite valleys is unique to monolayer TMDs.

In addition to the emergence of this new peak, the cross peaks *XT* and *TX* red-shift along the emission energy axis by ~5 meV for cross-circular polarization. This polarization-dependent shift is consistent with the formation of a five-particle charged biexciton arising from a bound state between the exciton and negative trion—an interpretation that is supported by theoretical calculations of the optical response derived using Liouville space pathways[40], as discussed below. We emphasize that signatures of the neutral and charged biexcitons observed here are most likely masked by the trion transition in emission (projection of the 2D spectrum onto the emission energy axis) in linear optical spectra. A comparison of slices from the co-circularly and cross-circularly polarized 2D spectra taken along the emission energy axis at the exciton excitation energy, shown in Fig. 3c and Fig. 3d, highlights the exciton, biexciton, and charged biexciton states. We note that the laser bandwidth is broad enough to search for bound states with binding energies as large as 80 meV. The absence of additional peaks at lower emission energies further supports our assignment of the new peak *XX* between the exciton and trion as the bound neutral biexciton.

Further insight into the origin of the biexciton states is obtained through calculations of the co- and cross-circularly polarized 2D spectra, which account for exciton-exciton, trion-trion, and exciton-trion interactions phenomenologically (see the Supplementary Information for details). The third-order polarization is evaluated for finite pulses in the rephasing time ordering using an energy level scheme consisting of singly and doubly excited manifolds of the exciton and trion states. The valley-specific polarization selection rules are explicitly considered here. A calculated 2D spectrum for co-circular excitation is shown in Fig. 4a. The spectrum features two diagonal and two off-diagonal peaks similar to the measured spectrum shown in Fig. 3a. For this polarization sequence, only the exciton and trion transitions in one valley are relevant. The diagonal population peaks are associated with the singly-excited excitons and trions. The off-diagonal coupling peaks appear un-shifted in the excitation and emission energies with respect to exciton and trion resonances and originate from coherent interactions between these two resonances. Due to the co-circular excitation scheme considered here and the valley-specific optical selection rules, many-body effects associated with interaction-induced shifts, or binding between multiple quasiparticles, will not contribute to the signal within the spectral bandwidth since only the singly excited manifolds are accessible.

The simulated spectrum for the case of cross-circular polarized excitation pulses is shown in Fig. 4b—the key features of which are consistent with the measurements. Most importantly, the doubly-excited state manifold is now accessible through the $g \to e \to f$ pathway (inset to Fig. 4b), where $e$ and $f$ represent one- and two-quasiparticle excitation manifolds (Suppl. Fig. 2). In the absence of Coulomb interactions between quasiparticles, the transitions $g \leftrightarrow e$ and $e \leftrightarrow f$ are quantum mechanically the same and cannot be distinguished. As a result, the quantum pathways of the singly- and doubly-excited states completely cancel and the nonlinear signal is zero. Therefore, the fact that a signal is observed for this polarization sequence and for each peak type—diagonal and off-diagonal ones—necessarily implies that many-body effects stemming from exciton-exciton, exciton-trion, and trion-trion interactions dominate the nonlinear optical response. The neutral biexciton peak *XX* and the charged biexciton peaks (*XT* and *TX*) are present in the calculated cross-polarization spectra with binding energies taken from the experiment in accordance with earlier calculations[15,16].

Interestingly, the weaker trion peak *T* compared to the exciton peak *X* for cross-circular polarization compared to the co-circular case in both measured and calculated spectra implies that trion-trion interactions are significantly weaker. Exciton-exciton interactions produce a tightly bound biexciton state, whereas the interaction shift for the trion ($\Delta_{TT}$) is estimated from a comparison of the exciton and trion relative amplitudes to be an order of magnitude smaller (~2 meV) likely due to weak localization at potential fluctuations and spatial separation of trions[41]. As a result, the quantum pathways associated with the singly- and doubly-excited states of the trion destructively interfere and lead to a small amplitude for the trion peak *T*. This interpretation is consistent with the slightly smaller exciton-trion (charged biexciton) interaction shift $\Delta_{XT}$ observed here compared to theoretical studies reported in the literature thus far, which assume complete delocalization of trions in the 2D plane[15,16]. Note that the position and relative intensity of the trion peak *T* are also influenced by effects such as the electron density in the doped sample, admixture of unwanted polarization components, or effects beyond $\chi^{(3)}$, which are not covered by the phenomenological model applied here.

In summary, we have observed unambiguous spectroscopic features of higher-order correlated states in monolayer MoSe$_2$, which we interpret as the neutral and charged biexcitons. A comparison of the nonlinear optical response under co- and cross-circular polarization conditions reveals both neutral and charged bound biexcitons with binding energies of ~20 meV and ~5 meV, respectively. These values are consistent with recent theoretical models predicting that the biexciton and trion have a similar electron-hole correlation function, which plays a critical role in their formation and dynamic optical response[15]. Characterization of the exciton-exciton, exciton-trion, and trion-trion binding energies presented here should assist future theoretical studies aiming to better understand many-body interactions in two-dimensional materials and help guide efforts searching for more exotic complexes including even higher order bound states[42] and polariton condensates[43].

## Methods

**Optical Two-Dimensional Coherent Spectroscopy (2DCS)**

40-fs pulses generated from a mode-locked Ti:Sapphire laser at a repetition rate of 80 MHz are split into a set of four phase-stabilized pulses using a system of nested Michelson interferometers. Three of the pulses are focused to a single ~30 μm spot on the sample. The coherent interaction of the pulses with the sample generates a four-wave mixing signal that is emitted in the wavevector phase-matched direction. The signal is recorded as the delay $t_1$ is stepped with interferometric precision. Subsequent Fourier transformation yields a rephasing one-quantum spectrum $S(\hbar\omega_1, t_2, \hbar\omega_3)$. Because the conjugated pulse $\mathcal{E}_1$ arrives at the sample first, the optical coherences evolve during $t_1$ with opposite phase as the coherences generated by $\mathcal{E}_3$ during $t_3$. As a result, the spectra are plotted with negative excitation energy. We hold $t_2 = 0$ fs to obtain maximum signal-to-noise; however, using a value greater than the pulse duration provides similar results at an overall smaller signal amplitude.

## Figure Legends

**Figure 1: Excitons, biexcitons, and trions in monolayer MoSe$_2$.** **(a)** Illustration of an exciton (*X*), trion (*T*), neutral biexciton (*XX*), and charged biexciton (*XT*) in monolayer MoSe$_2$. **(b)** Lowest-energy direct-gap transitions at the *K* and *K'* valleys associated with *A*-excitons are accessible using σ+ and σ- circularly polarized light, respectively. The valence band associated with the *B*-exciton is omitted for clarity. **(c)** Optical microscope image of the monolayer sample. **(d)**

Photoluminescence spectrum taken at 13 K featuring two peaks near ~1650 meV and ~1620 meV attributed to the exciton (*X*) and trion (*T*), respectively.

**Figure 2: Illustration of 2DCS of biexcitons.** **(a)** Box geometry used for the 2DCS experiments. Three pulses interact nonlinearly with the sample to generate a four-wave mixing signal that is collected in transmission and detected through heterodyne spectral interferometry. **(b)** Time ordering of the excitation pulses and detected signal. **(c)** The quantum pathway for accessing the bound biexciton (*XX*) for cross-circular polarization. Interactions between two excitons are modeled using a four-level system. The first pulse (σ+) creates an exciton coherence in the *K* valley during $t_1$. The second pulse (σ-) excites the exciton transition in the *K'* valley. The third pulse (σ+) drives the transition between biexciton and exciton states, which radiates as the four wave mixing signal.

**Figure 3: Polarization-resolved 2D coherent spectra revealing the biexciton resonances.** **(a)** Normalized 2D amplitude spectrum obtained using co-circular polarization of the excitation pulses and detected four-wave mixing signal. The spectrum features two diagonal peaks corresponding to the degenerate excitation and emission of the exciton (*X*) and trion (*T*) resonances and two off-diagonal peaks (*XT* and *TX*) corresponding to their coupling through Pauli blocking (saturation) nonlinearities. **(b)** Normalized 2D amplitude spectrum obtained using cross-circular polarization of the first/third fields (σ+) and second/signal fields (σ-). The additional peak (*XX*) is associated with the neutral bound biexciton, while the shift of the off-diagonal peaks to lower emission energy is associated with the charged bound biexciton. Slices along the emission energy axis at an excitation energy of 1648 meV are shown in **(c)** and **(d)** for co-circular and cross-circular polarization, respectively. The data are fit with a double and triple Gaussian functions (solid lines), respectively, with the individual fits to peaks *XT*, *XX*, and *X* indicated by the shaded regions.

**Figure 4: Calculated 2D spectra incorporating the singly- and doubly-excited state manifold.** Calculated 2D amplitude spectrum for **(a)** co-circular and **(b)** cross-circular polarization of the excitation pulses and detected signal. In the case of cross-circular polarization, both the singly and doubly excited states indicated by the inset are accessible.

**Acknowledgements** The spectroscopic experiments performed at UT-Austin were supported by NSF DMR-1306878 and NSF EFMA-1542747. X. Li also gratefully acknowledges the support from the Welch Foundation via grant F-1662 and a Humboldt fellowship, which facilitated the collaboration with TU-Berlin. P. Nagler, C. Schüller and T. Korn gratefully acknowledge technical assistance by S. Bange and financial support by the German Research foundation (DFG) via GRK 1570 and KO3612/1-1. J. F. Specht, M. Richter, and A. Knorr gratefully acknowledge discussions with Anke Zimmerman, GRK 1558, and financial support via Sonderforschungsbereich 787 (DFG).

**Author Contributions** G.M. and X.L. conceived the concept. K.H. led the experimental effort. All co-authors at the University of Texas ran the experiments, acquired the data, and analyzed the results. P.N., C.S. and T.K. provided the sample. J.F.S., M.R., and A.K. provided theoretical support. G.M., X.L., and K.H. wrote the manuscript. All authors discussed the results and commented on the manuscript at all stages.

**Figure 1: Excitons, biexcitons, and trions in monolayer MoSe₂.**

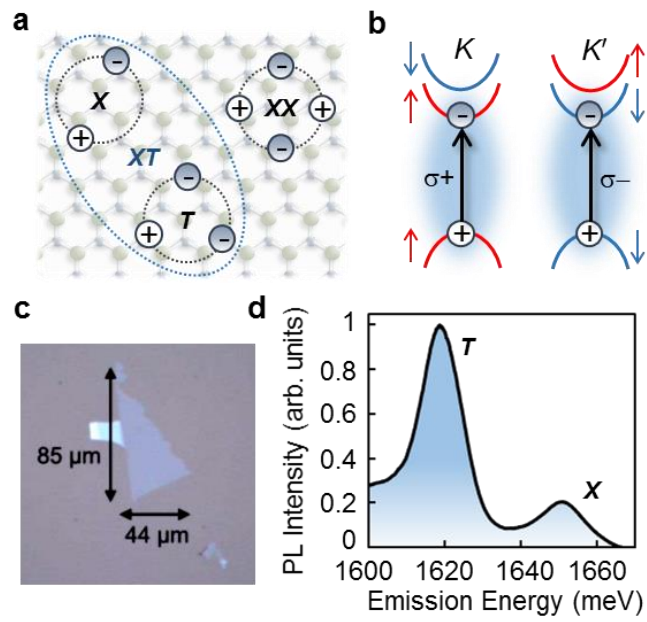

**Figure 2: Illustration of 2DCS of biexcitons.**

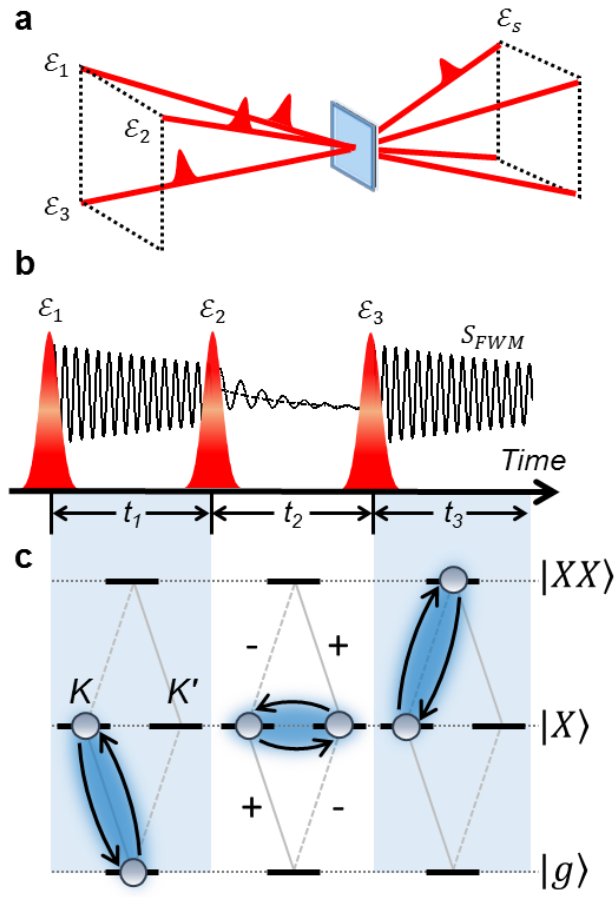

**Figure 3: Polarization-resolved 2D coherent spectra revealing the biexciton resonances.**

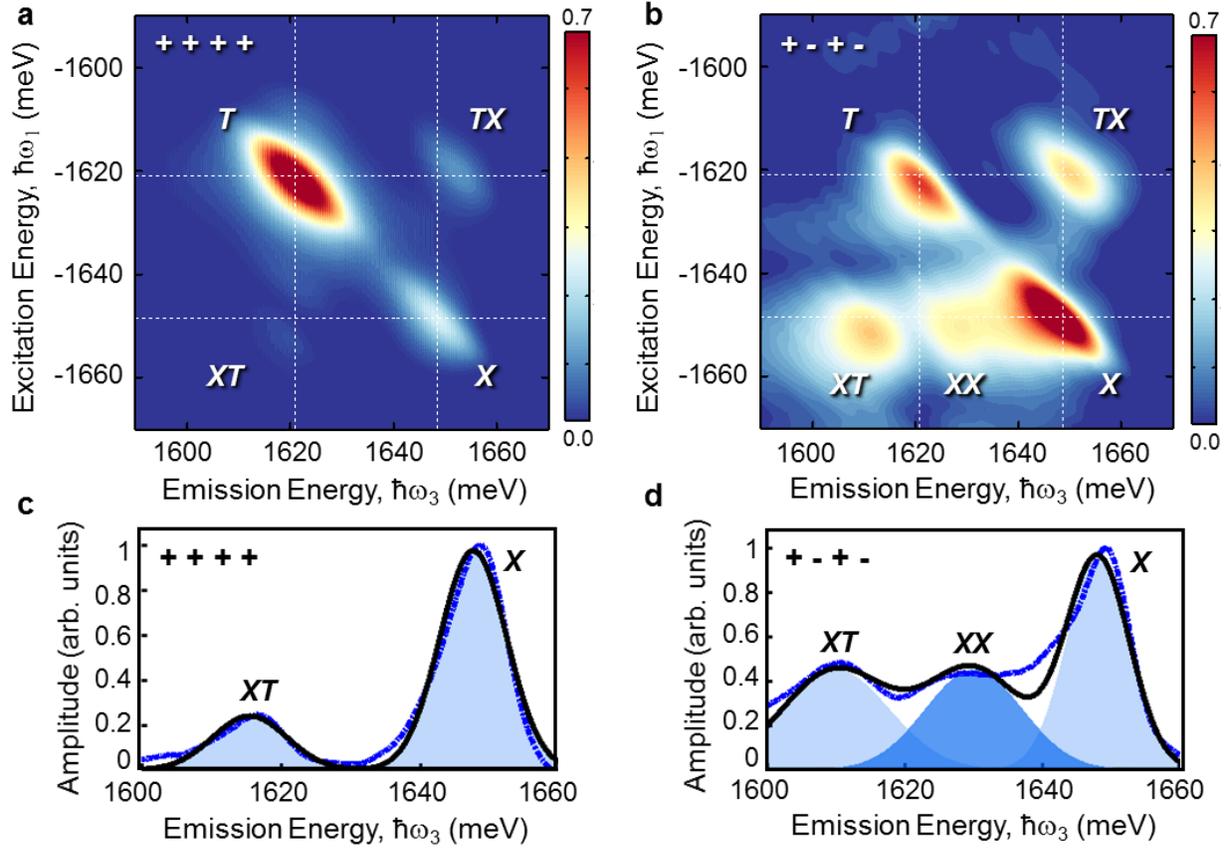

**Figure 4: Calculated 2D spectra incorporating the singly- and doubly-excited state manifold.**

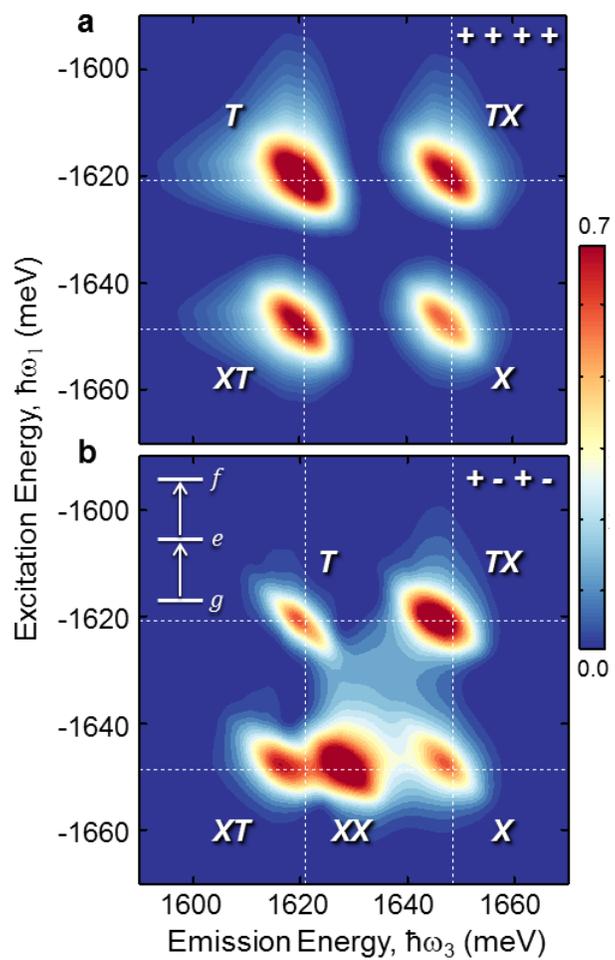